\documentclass[ASNA,twocolumn]{USG} %twocolumn
\usepackage{anyfontsize} %

\usepackage{colortbl}   % For coloring table lines
\usepackage{xcolor}     % For color definitions

\usepackage{steinmetz}
\graphicspath{{./images/}}
\specialissue{Morphologic Hard and Soft Tissue Changes following Tooth
Extraction in Different Maxilla and Mandible Anatomic Areas: Current
Knowledge, Prognostic Factors Analysis, and Clinical Management}

\articletype{ORIGINAL ARTICLE}%
\subarticletype{Particle Technology and Fluidization}

\received{20 January 2023}
\revised{21 January 2023} 
\accepted{22 January 2023}
\journal{AIChE Journal}
\volume{0}
\copyyear{2024}
\startpage{1}
\articledoi{10.1002/0000}

\begin{document}
\title{Point-and-Shoot Targeting for Efficient Neurosurgical Trajectory Planning in Mobile Augmented Reality}
%\subtitle{Subtitle}
%\transtitle{Guidelines for Establishing a Cytometry Laboratory}
%\subtranstitle{trans-subtitle}
\author[1]{Li Wah D'Angelo}
\author[1]{Pascal Spiegler}
\author[2]{Aristides Hadjinicolaou}[https://orcid.org/0000-0002-9926-2080]
\author[3,4]{Alexander G. Weil}[https://orcid.org/0000-0003-4162-6096]
\author[1]{Yiming Xiao}
[https://orcid.org/0000-0002-0962-3525]

%\author[7]{Uttara Chakraborty}[https://orcid.org/0000-0001-7037-2721]

%\author[8]{Kathleen T. Daniels}[https://orcid.org/0000-0001-7037-2721]

%\author[9]{Gelo de la Cruz}[https://orcid.org/0000-0001-7037-2721]

%\author[10]{Laura Ferrer-Font}[https://orcid.org/0000-0001-7037-2721]

%\author[11]{Andrew Filby}[https://orcid.org/0000-0001-7037-2721]

%\author[12]{David M. Gravano}[https://orcid.org/0000-0001-7037-2721]

%\author[13]{Michael D. Gregory}[https://orcid.org/0000-0001-7037-2721]

%\author[14]{Christopher Hall}[https://orcid.org/0000-0001-7037-2721]

%\author[15]{Christian Kukat}[https://orcid.org/0000-0001-7037-2721]

%\author[16]{André Mozes}[https://orcid.org/0000-0001-7037-2721]

%\author[17]{Diana Ordoñez-Rueda}[https://orcid.org/0000-0001-7037-2721]

%\author[18]{Eva Orlowski-Oliver}[https://orcid.org/0000-0001-7037-2721]

%\author[19]{Isabella Pesce}[https://orcid.org/0000-0001-7037-2721]

%\author[20]{Ziv Porat}[https://orcid.org/0000-0001-7037-2721]

%\author[21]{Nicole J. Poulton}[https://orcid.org/0000-0001-7037-2721]

%\author[22]{Kristen M. Reifel}[https://orcid.org/0000-0001-7037-2721]

%\author[23]{Aja M. Rieger}[https://orcid.org/0000-0001-7037-2721]

%\author[24]{Rachael T. C. Sheridan}[https://orcid.org/0000-0001-7037-2721]

%\author[25]{Gert Van Isterdael}[https://orcid.org/0000-0001-7037-2721]

%\author[14]{Rachael V. Walker}[https://orcid.org/0000-0001-7037-2721]

\authormark{D'Angelo \textsc{et al.}}
\titlemark{Point-and-Shoot Targeting for Efficient Neurosurgical Trajectory Planning in Mobile Augmented Reality}

\address[1]{\orgdiv{Department of Computer Science and Software Engineering, }\orgname{Concordia University, }%
\orgaddress{\state{Québec, }\country{Canada}}}

\address[2]{\orgdiv{Department of Pediatrics, Division of Neurology, }\orgname{Sainte-Justine University Hospital Center, }\orgaddress{\state{Québec, }\country{Canada}}}

\address[3]{\orgdiv{Department of Surgery, Division of Neurosurgery, }\orgname{Sainte-Justine University Hospital Center, }\orgaddress{\state{Québec, }\country{Canada}}}

\address[4]{\orgdiv{Department of Surgery, }\orgname{University of Montreal Hospital Center (CHUM), }\orgaddress{\state{Québec, }\country{Canada}}}

\corres{Yiming Xiao (\email{yiming.xiao@concordia.ca}) }

%\editor{\textbf{Academic Editor:} xxx ~|~ \textbf{Guest Editor:} xxxx}

%\presentaddress{This is sample for present address text this is sample for present address text.}

\fundingInfo{The study was supported by the Fonds de recherche du Qu\'ebec Nature and Technologies (FRQNT), Team Research Project Grant: 2022-PR296459 \cite{xiao2021frqnt}; Fonds de recherche du Qu\'ebec-Sant\'e Junior 1 Research Scholar program, and Parkinson Qu\'ebec \cite{xiao2024frq}.}


%\transkeywords{DEM | flexible cylindrical particle | internal friction angle | particle deformation | shear stress}

\abstract[ABSTRACT]{Deep brain stimulation (DBS) is an effective surgical procedure for drug-refractory Parkinson’s disease, and requires accurate planning of an electrode insertion trajectory to reach the surgical target while avoiding vital anatomies (e.g., blood vessels). With the need to navigate complex anatomy, conventional neuronavigation software that relies on 2D re-slicing of volumetric scans can be time-consuming and cumbersome to use. With higher flexibility and portability than common extended-reality headsets, mobile augmented reality (AR) allows for direct manipulation of 3D models of the patient’s anatomy, making it an excellent medium for surgical planning in the clinic. We propose a tablet-based mobile AR system to enable efficient neurosurgical trajectory planning for DBS. Specifically, we propose and compare two novel mobile AR ``point-and-shoot"-based methods (``single-line" vs. ``two-line") to place designated 3D mid-air points within anatomical mesh models. By adopting the preferred ``two-line" method to annotate the surgical target, we extended and evaluated the ``point-and-shoot" strategy to determine the optimal surgical path by interactively aligning the surgical target with an entry point. Through the user study with both lay users and clinicians, the proposed mobile AR prototype for DBS planning has a 73.30±11.45 system usability scale (SUS) rating, sub-millimeter surgical planning accuracy, and only $\sim$2 min planning time for a single trajectory. }

 \transabstract[transABSTRACT]{This is a generic template designed for use by multiple journals,
which includes several options for customization. Please
refer the author guidelines and author LaTeX manuscript preparation document for the journal to which you
are submitting in order to confirm that your manuscript will
comply with the journal’s requirements. Please replace this
text with your abstract. This is sample abstract text just for the template display purpose.}

%\abbr{5-FU, 5-fluorouracil; CFD, computational fluid dynamics; CH, channel; EFS, event-free survival; GBM, glioblastoma multiforme; OS, overall survival; PFS, progression-free survival; SD, standard deviation.}

%\contributed{Hao Zhang and Pengyue D. Guo contributed equally to this study.}

%\dedicated{Dedicated to Srivivasa Ramanujano on the occasion of his 125th birth anniversary.}

\copyright{This is an open access article under the terms of the \href{Creative Commons Attribution-NonCommercial}{Creative Commons Attribution-NonCommercial} License, which permits use, distribution and reproduction in any medium, provided the
original~work~is~properly cited and is not used for commercial purposes.
\\[5pt]
  ©  2024 The Author(s) \textit{AIChE Journal} published by Wiley Periodicals LLC on behalf of American Institute of Chemical Engineers.}

%\openaccessstatement

\maketitle
%\afterpage{\aftergroup\restoregeometry}

%\onecolumn

\section{Introduction}\label{sec1}

%% \section{Introduction} %for journal use above \firstsection{..} instead
As the second most frequent neurodegenerative disorder, Parkinson’s disease (PD) primarily presents motor dysfunctions, while non-motor symptoms, including psychiatric problems and sleep disturbance, can also occur. Deep brain stimulation (DBS) is an effective surgical treatment for drug-resistant PD to relieve its motor symptoms, such as tremor, body rigidity, and bradykinesia. It achieves the therapeutic goal by delivering chronic electric stimulation to a designated deep brain nucleus (a small cluster of neurons) through an implanted electrode. With the common surgical target of the subthalamic nucleus (STN), a small lentil-sized nucleus in the midbrain, safety and treatment outcome for the procedure greatly depend on the quality of pre-surgical planning, where an electrode insertion trajectory is designed to accurately target the therapeutic location while avoiding other crucial structures (e.g., blood vessels, ventricles, and sulci) that can lead to complications. When a sub-optimal surgical trajectory is deployed, additional trial-and-error adjustments in the operating room (OR) are often required, resulting in prolonged surgical time, hemorrhage, and other adverse side-effects. Due to the need to navigate different neuroanatomical structures, multiple MRI contrasts that best visualize different anatomies are necessary for optimal surgical planning. However, conventional neuronavigation software for surgical planning primarily relies on 2D display of re-sliced volumetric imaging data to visualize the spatial relationship of complex brain anatomy (especially blood vessels) from multi-contrast scan fusion. This poses great challenges to the clinician’s cognitive load and spatial memory skill when it comes to determining a viable neurosurgical trajectory \cite{Beriault2012MultiModal}, resulting in lengthy planning time ($\sim$ 40 min \cite{Beriault2011}). Although automatic trajectory planning algorithms \cite{Zelmann2015Improving,Beriault2012MultiModal} leverage different cost criteria to optimize surgical trajectories (often with multiple viable solutions), such as the distances to the blood vessels and ventricles, they still require ground truth data to help determine the weights for each criterion in the final cost function and could still benefit from interactive software to inspect and revise the surgical plans. \textbf{Therefore, an efficient and intuitive manual trajectory planning method is in demand.}

Extended-reality (XR) technology offers intuitive visualization and interaction of complex patient-specific volumetric data. Recent technological progress in wearable XR devices, such as the \textit{Meta Quest}, Microsoft HoloLens, and \textit{Apple Vision Pro} headsets has led to development of software applications that leverage these tools for medical training/education and surgical planning/monitoring. Besides major efforts that were dedicated for medical training/education \cite{tene2024virtual,pinter2020slicervr,Sonia}, a few solutions have been proposed for the planning of various surgical interventions. For example, with virtual reality (VR), Hellum et al. \cite{Hellum2022Voodoo} proposed to use voodoo doll annotation and an eye-tracking-based trajectory design for efficient DBS trajectory planning. In thoracic surgery, Hairirpoush et al. \cite{Harirpoush2024} designed a VR-based system to optimize trocar placement and surgical tool trajectory for reducing surgical complications and time. More recently, Spiegler et al. \cite{Spiegler2025} devised and assessed a user-friendly VR-based SEEG electrode trajectory planning tool. \textbf{On the other hand, prior work also demonstrates that augmented reality (AR) can improve spatial awareness and procedural understanding in clinical contexts, such as for surgical planning and navigation}. For instance, Fotouhi et al. \cite{fotouhi2020development} developed a spatiotemporal-aware AR system with Microsoft HoloLens for C-arm-based orthopedic interventions that improves visualization of anatomy and surgical tools. Similarly, Moglia et al. \cite{moglia2023mixed} integrated mixed reality with deep learning-based anatomical segmentation to support multi-modal visualization and interaction in knee osteotomy planning. In neurosurgery, Kunz et al. \cite{kunz2021autonomous} leveraged AR-based visualization for intraoperative guidance of ventricular puncturing (lower accuracy requirement than DBS, in the range of $\sim$6 mm). Furthermore, recent reviews highlight the broader potential of AR for improving surgical precision across specialties, such as spine surgery \cite{de2024integrating}. Compared with high-end VR and AR head-mounted displays typically used in the previous literature, mobile AR that leverages smartphones and tablets can be more cost-effective, flexible, and portable. In addition, these mobile devices are increasingly being integrated into digital patient record communication (e.g., note-taking, radiological data transfer and inspection, etc.), making them readily available in the clinical environment. With mobile augmented reality (AR), Mehta et al. \cite{Mehta2024} developed ARCollab to enable multi-user interactive cardiovascular surgical planning. Asadi et al. \cite{Asadi2025} built iSurgARy, a mobile AR system to facilitate the guidance of ventriculostomy procedure in resource-limited settings. \textbf{However, most of the existing surgical XR systems still primarily focus on visualization and navigation, leaving the design of robust spatial input methods with metric-based feedback for surgical trajectory specification under-explored, especially in mobile AR}.

In contrast to head-mounted XR devices, mobile AR faces the challenges of missing stereo-vision and tracked controllers for 3D mid-air point annotation within virtual anatomical models \cite{Hellum2023}, and thus requires tailored solutions for virtual object manipulation and annotation. A comprehensive survey by Goh et al. \cite{goh20193d} categorizes common interaction paradigms for handheld/mobile AR into touch-based manipulation, screen-centered ray casting, and device-motion–based interaction, highlighting the challenges of controlling depth and achieving precise 3D placement using a 2D touchscreen interface. To mitigate depth ambiguity, Polvi et al. \cite{polvi2016slidar} proposed SlidAR, which constrains object placement to detected surfaces reconstructed from SLAM, allowing users to position objects by sliding them along real-world geometry. More recent work by Song et al. \cite{song2023leveraging} leverages tablet motion tracking to combine device movement with touch interaction, enabling intuitive viewpoint-driven manipulation for operating room scene annotation tasks, such as bounding box labeling. Similarly, Grandi et al. \cite{grandi2018design} explored collaborative interaction techniques that allow multiple users to manipulate virtual objects in shared handheld AR environments. Despite these advances, existing methods are primarily designed for surface-based placement or general object manipulation. \textbf{So far, precisely specifying points inside complex anatomical models for surgical planning remains challenging due to depth ambiguity and limited control over ray direction.} This motivates further development of robust interaction techniques, specifically for accurate mid-air point placement and trajectory specification in mobile AR environments (e.g., tablet AR).

To address the aforementioned challenges, we propose a novel mobile AR system (see Fig. 1), which combines multi-view exploration and touch-based interaction, and bridges general-purpose mobile AR input techniques and the precision requirements of neurosurgical planning. Specifically, our work has three major contributions. \textit{First}, we propose two novel “point-and-shoot" mobile AR interaction techniques (“Single-line" and “Two-line") for accurate mid-air surgical target annotation within anatomical mesh models for the first time. \textit{Second}, building on the preferred two-line method, we further extend the “point-and-shoot" strategy for an efficient, metric feedback-based trajectory planning technique, which reformulates surgical path specification as an intuitive spatial alignment task between entry and surgical target points. \textit{Lastly}, we systematically evaluate the designed surgical target annotation methods with six different mesh models and the full system with both lay users and clinicians. While demonstrated with the DBS procedure as an example, our system is also extendable to other surgical planning applications.

\begin{figure*}
    \centering
    \includegraphics[width=1\linewidth]{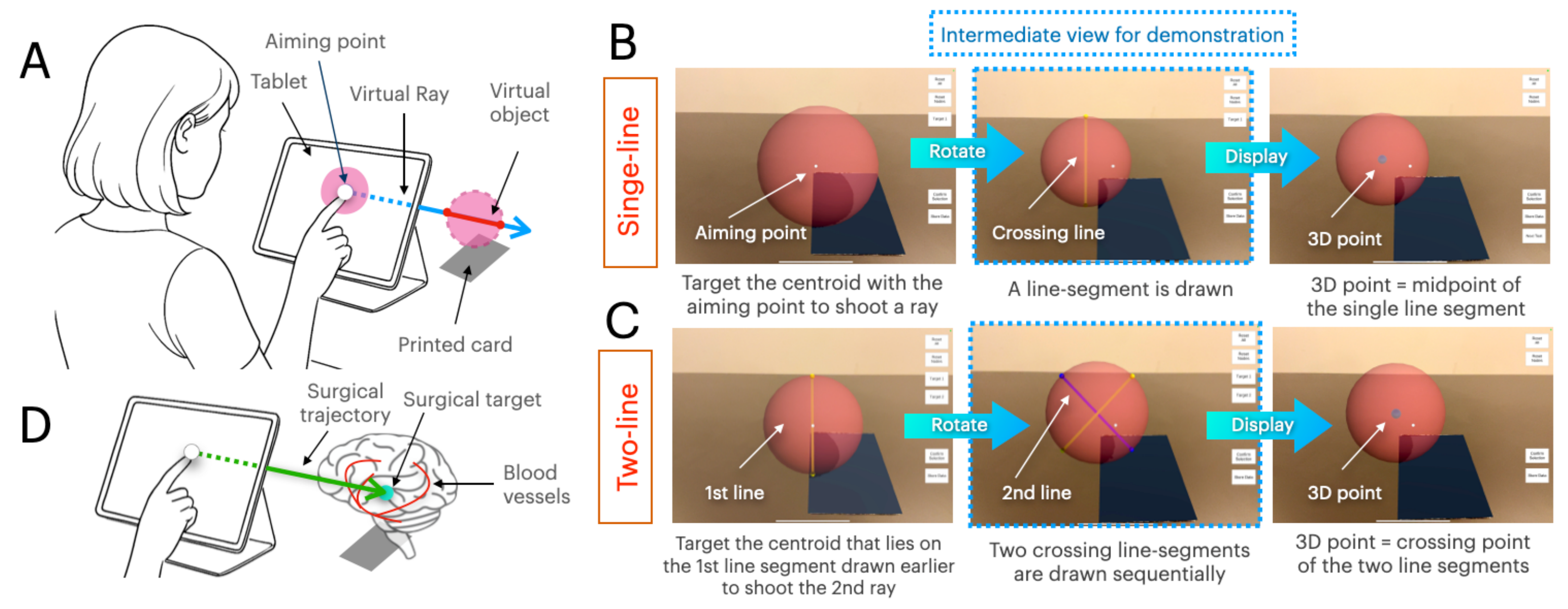}
    \caption{Demonstration of the proposed tablet-based mobile augmented reality (AR) system with the ``Point-and-Shoot" strategy for \textbf{neurosurgical target (3D point) placement} and \textbf{trajectory design}. A. Illustration of the user interaction setup for surgical target (3D point) placement. B. ``Single-line" strategy for surgical target placement, where the 3D point is placed as the center of the intercepting line between the 3D mesh and the projected ray from the tablet's aiming point. C. ``Two-line" strategy for surgical target placement, where the 3D point is placed at the crossing point of two rays sequentially generated from the aiming point within the mesh model. \textit{Note that to better demonstrate the projected line segments from ray-shooting, we show the rotated views of the object before the display of the final 3D point placement, but they are not necessary in real practice.} D. ``Point-and-shoot" based neurosurgical trajectory planning with surgical target aiming and blood vessel avoidance.}
    \label{fig:shapes}
\end{figure*}

 \section{Materials and Methods}\label{sec2}

\noindent
To establish the proposed mobile AR system, we conducted a two-phase study, with the first phase investigating the suitable interaction technique for 3D mid-air point placement from two candidates and the second integrating it into a full neurosurgical trajectory planning workflow. The study was approved by the local Research Ethics Committee of our research institute.

\subsection{Digital model generation}

For our study, we constructed two sets of digital models. The first set was used for comparing 3D point placement techniques with mobile AR, and includes six different artificial and neuroanatomical mesh models (see Fig. 2). While three artificial shapes were made with Unity and Blender with their physical dimensions matching those of typical brain nuclei, the lateral ventricle and hippocampus models were extracted from real brain MRIs \cite{Xiao2019}, and the aneurysm model was obtained from the IntrA dataset \cite{yang2020intra}. Specifically, the simple sphere of Shape I has a diameter of 10 mm, the small sphere in Shape II has a radius of 8 mm, and Shape III stretches $\sim$18 mm. In terms of the anatomical models, Shape IV (lateral ventricle) stretches 75 mm in the left-right direction and 95 mm in the anterio-posterior direction, the aneurysm in Shape V has a $\sim$3-mm diameter, and Shape VI (hippocampus) has a $\sim$50-mm length. These physical dimensions provide the context for the measurement of 3D point placement errors.

For the second phase to assess the full DBS trajectory planning method, we constructed a digital brain model based on anatomical segmentations of 3T brain MRI scans from a healthy subject. Upon informed consent, the subject was scanned with T1-weighted and T2-weighted MRIs, as well as time-of-flight (TOF) MR angiography (MRA) using a 3T MRI scanner. Brain surface, bilateral midbrain nuclei (red nucleus, substantia nigra, and subthalamic nucleus) \cite{Xiao2019}, and cerebral vasculatures were segmented from the MRIs. Specifically, the blood vessels were segmented using the Frangi vesselness filter~\cite{Frangi} and then refined manually by the senior author using ITK-SNAP (http://itksnap.org). The midbrain nuclei were obtained by warping the ICBM152 atlas to the subject's T2w MRI. The brain surface was extracted using the BEaST algorithm ~\cite{BEAST}. In the constructed AR view, the vasculature, sulci, and lateral ventricle are rendered with red, blue, and yellow mesh models to allow maximum visual contrast for surgical planning, while the brain surface is shown as a semi-transparent surface to show the overall brain shape.

% \vspace{10pt}

% {\setlength\arrayrulewidth{5pt}
% \arrayrulecolor{gray}
% \begin{tabular}{|p{230pt}}
% Lorem ipsum dolor sit amet, consectetur adipiscing elit. Morbi id hendrerit nunc. Sed scelerisque lacus vitae erat eleifend eleifend. Donec eros mi, placerat in porta eleifend, placerat a urna. Pellentesque venenatis neque non turpis convallis vehicula. Aliquam aliquet ultricies tincidunt.
% \end{tabular}}

\begin{figure}
    \centering
    \includegraphics[width=1\linewidth]{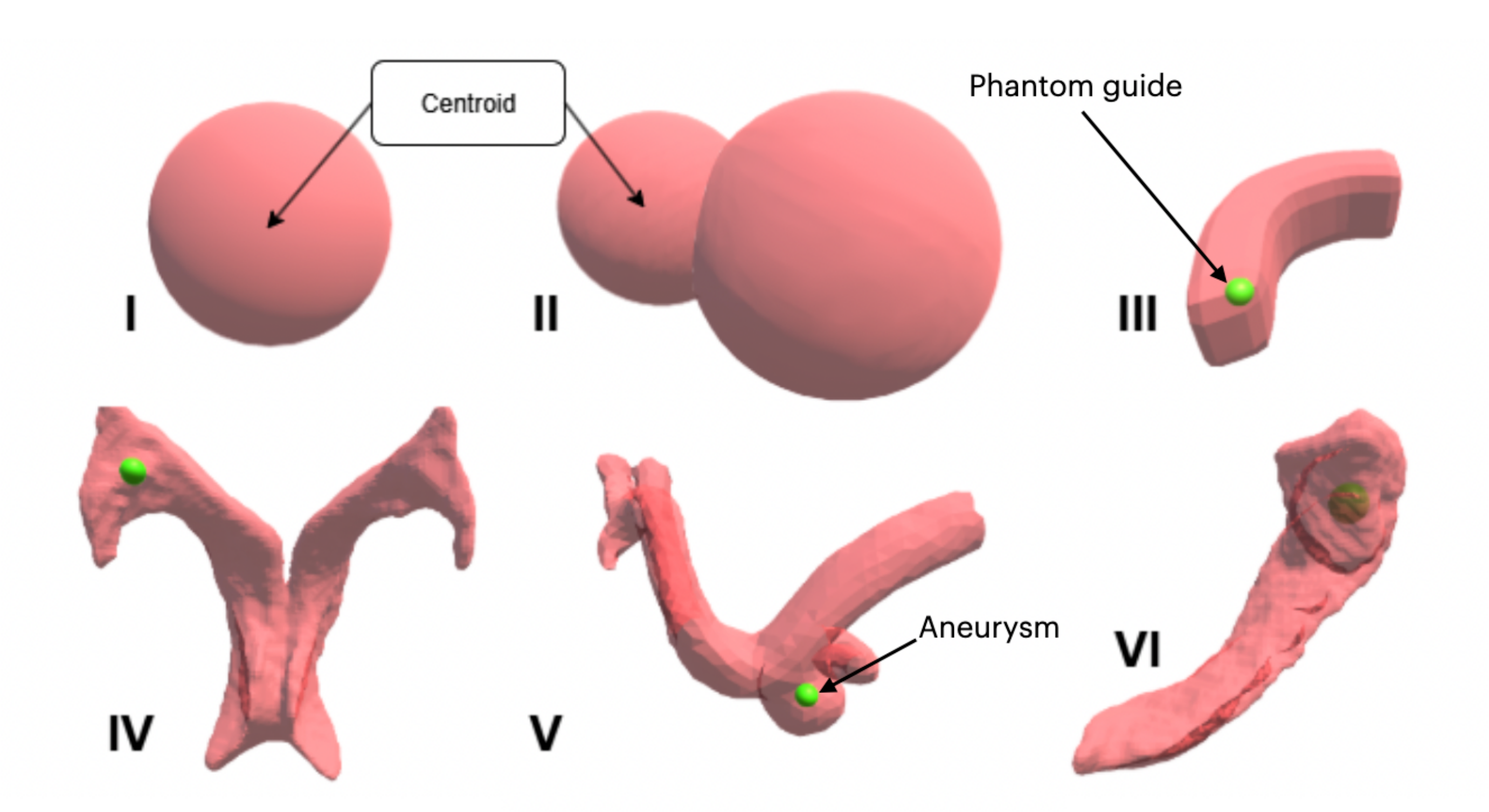}
    \caption{Illustration of six different shape models for testing 3D point placement methods. While Models I-III are artificial shapes, Models IV, V, and VI are the lateral ventricle, cerebral aneurysm, and left hippocampus.}
    \label{fig:shapes}
\end{figure}

\subsection{Mobile AR system setup}
%basic setup for the card-based AR setup
Our tablet-based AR system was developed using the Unity 3D game engine (Version 2022.3.3f1) while the Vuforia Engine (Version 10.27.3) was used to implement the AR-based digital model display, allowing printed cards with design patterns to be spatially tracked with the front camera of the tablet and augment the virtual models above it. To avoid distraction from the printed pattern on the tracker card when viewing the virtual objects, we overlay a virtual black card of the same size on top of the real one. For the experiment, we used a 10.9-inch iPad (10th generation) mounted on a tablet holder stand. Finger gesture-based spatial manipulation of the virtual model was implemented for two-finger-based uniform model scaling and translation (left-right \& up-down), and one-finger-based model rotation. Alternatively, object rotation and translation can also be achieved by moving the printed card and/or the tablet device.

\subsection{Ray shooting-based 3D point placement}

In surgical planning, 3D anatomical point annotation is a fundamental task to mark surgical targets and perform measurement. Often, these points need to be within an anatomical volume rather than on its surface, which are difficult to annotate with conventional raycast-and-collide methods in XR, particularly for tablet-based mobile AR. To address this, we introduce a ``point-and-shoot" strategy as demonstrated in Fig. 1. To place a point in the AR objects, we raycast a virtual line to the displayed virtual object from the aiming point at the center of the iPad, shown as a white dot on the screen (see Fig. 1B). When the ray hits the object, it will mark an entry point, an exit point, and a line segment that connects the entry point to the exit so that a desired 3D point can be placed along this line segment. To further localize the spatial point, we proposed and compared two approaches: ``single-line" vs. ``two-line". 

\begin{figure}
    \centering
    \includegraphics[width=1\linewidth]{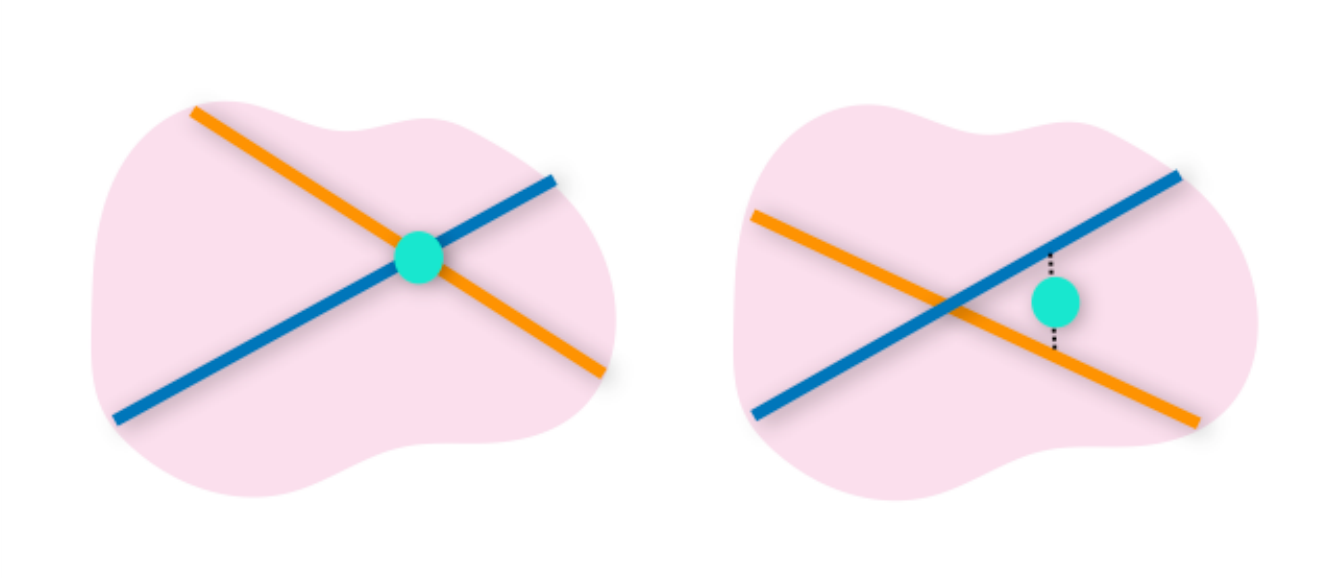}
    \caption{Demonstration of mid-air point annotation within an anatomical mesh model with the ``Two-Line" method. Left: with precise aiming, the 3D point is placed at the cross of two line segments; Right: when the aiming is less precise, the 3D point is placed as the mid-point of the line segment that is the shortest distance between the two lines.}
    \label{fig:linedemo}

    \vspace{-0.3cm}
\end{figure}

In the ``\textbf{single-line}" approach (Fig. 1B), the user translates/rotates the virtual object with finger gestures (and/or via tablet motion and repositioning of the printed card) so that the aiming point at the center of the screen targets the desired location in the view. Then, upon tapping the confirmation button, a ray is shot from the aiming point through the object and a 3D point is annotated at the mid-point of the line segment enclosed within the object along the casted ray. For example, to annotate an arbitrary point within a 3D object, the user needs to strategically orient it so that the midpoint of the resulting line segment between the front surface facing the user and the opposite side aligns with the desired point location. While this can be straightforward for tagging centroids of symmetric shapes (e.g., spheres), for points within irregular shapes, the user will need higher spatial cognition skill to select the optimal viewing angle for the task with this technique.

As an extension to the ``one-line" technique, the ``\textbf{two-line}" approach (Fig. 1C) requires the user to shoot two rays sequentially through the object, where the desired 3D point will be placed at the intersection of the resulting two line segments enclosed within the mesh object. However, due to individual variability in finger gesture precision, the casted two rays may not always intersect perfectly. In this case, the 3D point will be placed at the midpoint of the line segment that represents the shortest distance between them (Fig. \ref{fig:linedemo}). To allow the optimal result, the two casted rays should be as perpendicular as possible to each other, as this helps ensure sufficient viewing angle exploration to compensate for the lack of stereo-vision in mobile AR, particularly for complex shapes.

\begin{figure*}[t]
    \centering
    \includegraphics[width=\linewidth]{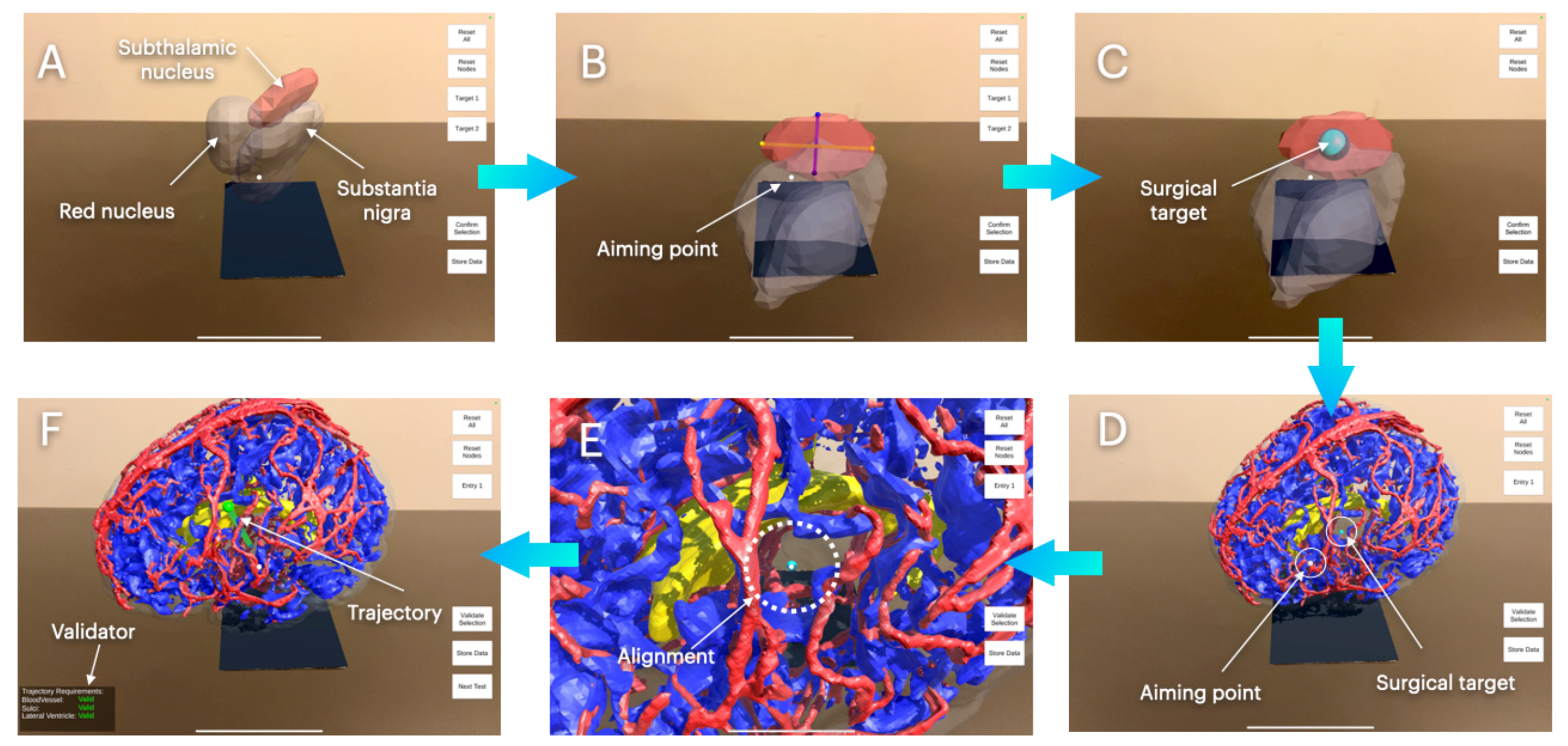}
    \caption{Demonstration of workflow of DBS trajectory planning for the left subthalamic nucleus (STN). A to C: Surgical target annotation with the ``two-line" crossing ray method; D. The coordinate of the determined surgical target is transferred to the full brain model; E. The user rotates the brain model with finger gesture to align the aiming point and the surgical target while avoiding the blood vessel (red), sulci (blue), and ventricle (yellow); F. Generate the surgical trajectory after validating the distances to the nearest blood vessel, sulci, and ventricle.}
    \label{fig:trajectory}
\end{figure*}

\subsection{DBS trajectory planning}

With the same ``point-and-shoot" paradigm, we designed a tablet-based mobile AR system to perform bilateral DBS trajectory planning, which is separated into two steps, as demonstrated in Fig. \ref{fig:trajectory}. With the system, the user will design one trajectory for each side of the brain at a time (left side first).

\noindent
\textbf{Step 1:} The user is presented with the anatomical model of the subthalamic nucleus (STN) (Fig. \ref{fig:trajectory}A) and must place a surgical target point at its centroid using the winning approach between the two introduced methods as determined through the Phase 1 user study (i.e., ``two-line" as shown in Fig. \ref{fig:trajectory}B). The adjacent red nucleus and the substantia nigra in the same brain hemisphere are also visualized to provide the spatial and anatomical context for the STN. Once the surgical target point is placed and confirmed (Fig. \ref{fig:trajectory}C), its coordinates are recorded for Step 2. 

%The scale of the surgical target is large such that it is the same size proportionally as in Step 2.

\noindent
\textbf{Step 2:} The user is presented with a brain model containing the brain surface, blood vessels, sulci, lateral ventricles, and the surgical target point (Fig. \ref{fig:trajectory}D). The surgical target point determined in Step 1 is mapped to the corresponding location in the full brain model and is shown as a bright cyan sphere. The user is tasked with determining a suitable entry point on the brain surface such that the trajectory connecting the entry point to the surgical target has enough clearance ($>$ 2 mm distance margin) from the blood vessels, sulci, and lateral ventricles. To do so, the user will use finger gestures to reorient the brain model so that the surgical target is 1) visually exposed through a clear path free from the vital structures and 2) aligned with the aiming point on the tablet screen (Fig. \ref{fig:trajectory}E). Upon confirming the choice, the aiming point will shoot a ray to intercept with the brain surface to obtain the entry point, and hence the surgical trajectory. After confirming the trajectory, the system will validate if the designed trajectory is far enough from the other vital anatomical structures (Fig. \ref{fig:trajectory}F). If the trajectory is within 2 mm to any of them, the current surgical plan will be cleared and the user will repeat Step 2 to re-design the trajectory. Note that as the anatomical mesh models were obtained from the segmentation of the MRI scans, their coordinates are consistent with each other. Upon completion of the surgical plan, our system records the coordinates of the surgical target and trajectory entry point. These data can then be directly transferred to the intra-operative neuronavigation system to guide the intervention.

\subsection{User Study Design and System Assessment}
We conducted a two-phase user study, with Phase I comparing the two proposed 3D point placement techniques (``single-line" vs. ``two-line"), and Phase II evaluating the full DBS trajectory planning workflow with the winning point placement technique for surgical target annotation.

In \textbf{Phase I}, the participants will be presented with the six 3D models as shown in Fig. \ref{fig:shapes}, one at a time, and must place the target point as close as possible to their centroid or a designated point that is indicated as a ``phantom guide". For Shapes I and II, the user must approximate the centroid of the only sphere and the partially occluded small sphere, respectively. For Shapes III-VI, a phantom guide indicates where the target point is meant to be placed. For the anatomical shapes (IV-VI), these points represent potential surgical targets of interest in various procedures that require neurosurgical trajectory planning, such as the horn of the lateral ventricle for ventriculostomy, centroid of an aneurysm for cerebrovascular clipping, and the head of the hippocampus for SEEG implantation. For each 3D point placement technique, the user will complete the point annotation for all six objects in the same order (Shape I to VI) while the order of the point placement technique to be compared is randomized for all the participants to mitigate the ordering effect. The task completion time (in seconds), error of point annotation (in mm), and the number of attempts was recorded using the software system for each shape annotation and point placement method. In addition, the NASA Task Load Index (TLX) questionnaire \cite{NASATLX} was given to the participants to examine the task load for both techniques. Finally, a freeform user questionnaire was presented to ask the participants' preferences of the point placement technique, elaborate their NASA-TLX ratings, and provide suggestions of system improvements. We hypothesize that \textbf{\textit{the ``two-line" technique is superior to the ``single-line" in terms of task load, speed, robustness, and accuracy}}. This hypothesis was validated using the quantitative and semi-quantitative metrics and corroborated by the participants' selections. The winning technique will be incorporated in the full DBS planning for surgical target point annotation in Phase II.

In \textbf{Phase II}, the participants will assess the full workflow of the DBS trajectory planning by designing the trajectory for the left and right STN sequentially on the same brain. The procedure is detailed in Section II.D, and as revealed by the results of Phase I, the ``two-line" method is rated as the better choice for surgical target annotation. For all participants, we recorded the task completion time (in seconds), targeting error (in mm), and number of attempts for STN annotation, as well as the task completion time, number of attempts, and the final surgical trajectory's distances to the vital structures (blood vessel, ventricle, and sulci, in mm) for the trajectory planning step. Furthermore, we present the NASA-TLX and system usability scale (SUS) \cite{SUS68} questionnaires to assess the task load and usability of the proposed system. Finally, similar to Phase I, we invite the participants to elaborate their ratings in further detail and offer suggestions for future improvements.

Upon informed consent, for Phase I study, we recruited a total of 20 non-clinician participants (age = 25.9 $\pm$ 7.7 yo, 10 females) and five clinicians (1 neurology fellow and 4 neurosurgical residents/fellow, age = 28.8 $\pm$ 2.2 yo, 3 females), who work in the domain of functional neurosurgery, including DBS. To help contextualize the user study results, we also inquired the participants' familiarity with augmented reality and neuroanatomy in a 1-5 Likert scale (1 = Unfamiliar \& 5 = Familiar). Overall, the non-clinicians rated 3.3$\pm$1.2 and 2.3$\pm$1.3 for the familiarity level with AR and neuroanatomy, respectively. On the other hand, the clinicians gave their AR and neuroanatomy familiarity levels as 2.6$\pm$1.5 and 4.8$\pm$0.5, respectively. In Phase II, with informed consent, we invited the same participants from Phase I to test the full neurosurgical trajectory planning system, and 17 non-clinicians (age = 24.1 $\pm$ 2.8 yo, 8 females) and the same five clinicians participated in the study. Before the experiments started, for each participant, they were given a brief PowerPoint presentation regarding the context of the study, basic knowledge about deep brain stimulation and neuroanatomy, and a tutorial on how to navigate the proposed user interface. Upon completing the experiments, the participants were given a user questionnaire to help evaluate the software system as described earlier.

\subsection{Study Result Analysis}
To analyze the NASA-TLX score in Phase I, we normalize each sub-score to the range of [0,100], and the total NASA-TLX is computed as the average of all six normalized sub-scores. For the SUS questionnaire, the converted scores of ten questions are summed, and multiplied by 2.5, resulting in a maximum SUS score of 100. A software system that receives a SUS score above 68 indicates good usability \cite{SUS68}. For the SUS and all other acquired quantitative measures, we conducted the Lilliefors test to verify their normality in order to ensure the appropriate statistical tests for analysis. From the tests, we confirmed that the SUS scores across the participants followed a normal distribution, allowing the use of t-tests. On the other hand, all the remaining quantitative measures, except for the task completion times for point placements in the Ball (Shape I) and the point placement errors with the ``two-line" method for the Double Ball (Shape II) from Phase I failed to exhibit normal distributions, thus warranting non-parametric statistical tests instead. Therefore, for the ray shooting-based 3D point placement task (Phase I), Kruskal-Wallis tests were used to assess differences between two different techniques (``single-line" vs. ``two-line") in NASA-TLX scores, task completion time, point placement errors, and number of placement attempts. In the full DBS trajectory planning task (Phase II), the SUS scores were compared to the standard usability benchmark of 68 using a one-sample t-test. Note that a threshold of $p < 0.05$ was applied for all statistical analyses to confirm the significance of observed trends. As we only recruited 5 clinicians in our user study, we didn't perform statistical tests to compare group-wise differences between clinicians and non-clinicians due to limited statistical power. However, we report their metrics separately besides the results from all participants for the Phase II study. For the qualitative feedback from the questionnaires, we summarize the insights based on the frequencies of keywords used by the participants.

%\vspace{-0.4cm}

\section{Results}\label{sec3}

\subsection{3D point placement}

\begin{figure}
    \centering
    \includegraphics[width=\linewidth]{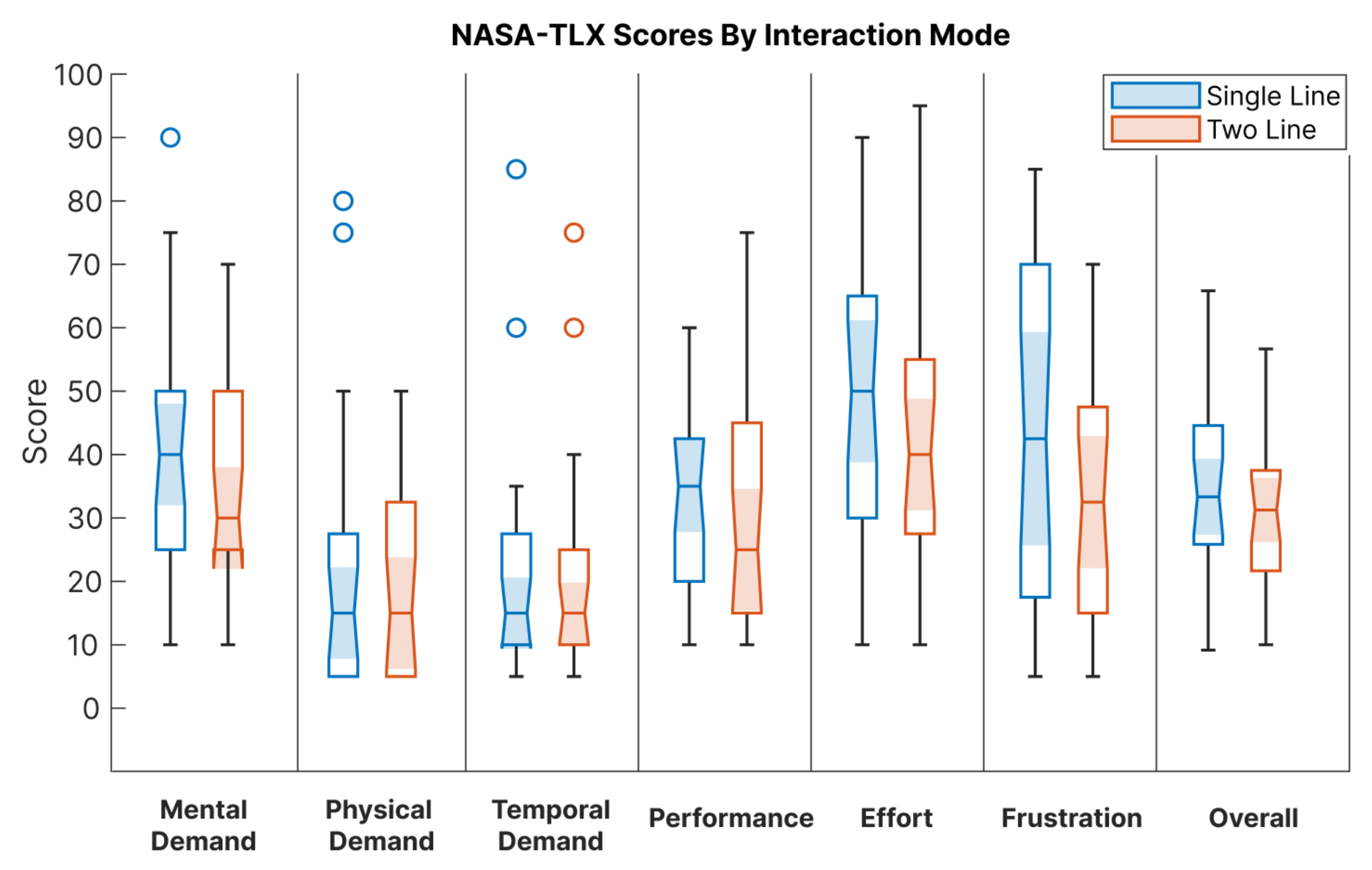}
    \caption{Boxplot of NASA-TLX scores for 3D point placement with ``single-line" and ``two-line" methods.}
    \label{fig:NASA-Point}
\end{figure}

For the Phase I study, when evaluating \textbf{NASA-TLX} (see Fig. \ref{fig:NASA-Point}), the ``two-line" method showed lower average task load than the ``single-line" counterpart on six out of seven sub-scores, including Mental Demand (35.83 ± 16.85 vs. 39.58 ± 20.58), Physical Demand (21.04 ± 17.13 vs. 22.08 ± 22.11), Performance (30.21±18.38 vs. 33.33 ± 14.87), Effort (40.62 ± 19.91 vs. 47.92 ± 20.74), Frustration (32.08 ± 19.33 vs. 43.12 ± 28.12), and Overall task load (30.17 ± 12.67 vs. 34.41 ± 15.39). The exception was Temporal Demand, which was slightly higher for the ``two-line" method (21.25 ± 17.65 vs. 20.42 ± 18.93). While none of these differences were statistically significant ($p > 0.05$), the consistent pattern suggests a slight perceived benefit of the ``two-line" approach, despite increased temporal demand. The objective quantitative metrics in Table \ref{tab:comparison_metrics_full} further show that the ``two-line" method significantly lengthens completion time only for the simpler Shape I: single ball target (125.79 ± 88.79 s vs. 74.97 ± 53.49 s, $p = 0.029$) and Shape II: double ball target (147.86 ± 93.61 s vs. 89.75 ± 88.59 s, $p = 0.0064$), but not for the complex ones. Despite this, the ``two-line" method resulted in significantly less repeated attempts (a measure of robustness) than the ``single-line" method for all shapes other than Shapes I and II. In addition, the targeting errors are on average higher for the ``single-line" method for the more complex Shapes III to VI, despite a statistically significant lower error for the ``single-line'' method on the simple Double Ball target (Shape II). However, this difference was only observed for this simple artificial shape and did not persist for the more complex and clinically relevant anatomical targets. These results demonstrate that the ``two-line" approach enhances mid-air 3D point placement accuracy without requiring more time on more complex, non‐spherical targets. \textit{Based on the observed experimental results, we concluded that the ``two-line" method is better than the ``single-line" approach.} This was further corroborated by the stated preference from almost all participants in their questionnaires.

% All other differences between methods did not reach significance.

\begin{table}[ht]
\centering
\caption{Mean±SD for Single-Line vs.\ Two-Line placement across all targets and metrics. The lower mean values are marked in \textbf{bold}. Asterisks (*) indicate statistically significant differences between the Single-Line and Two-Line methods (Kruskal–Wallis test, $p < 0.05$).}
\label{tab:comparison_metrics_full}
\setlength{\tabcolsep}{2pt}
\begin{tabular}{llcc}
\toprule
\textbf{Target}      & \textbf{Metric}         & \textbf{Single-Line}      & \textbf{Two-Line}         \\
\midrule
Ball                  & Time (s)                & $\mathbf{74.97 \pm 53.49}^{*}$         & $125.79 \pm 88.79^{*}$        \\
(Shape I)                      & Error (mm)                & $\mathbf{0.24 \pm 0.17}$       & $0.28 \pm 0.19$       \\
                      & Attempts               & $1.8 \pm 1.7$           & $\mathbf{1.4 \pm 0.8}$           \\
\addlinespace
Double Ball           & Time (s)                & $\mathbf{89.75 \pm 88.59}^{*}$         & $147.86 \pm 93.61^{*}$        \\
(Shape II)                      & Error (mm)                & $\mathbf{0.35 \pm 0.15}^{*}$       & $0.51 \pm 0.23^{*}$       \\
                      & Attempts               & $1.6 \pm 0.9$           & $\mathbf{1.4 \pm 0.7}$           \\
\addlinespace
Curved Shape          & Time (s)                & $147.39 \pm 120.55$       & $\mathbf{120.14 \pm 71.36}$        \\
(Shape III)                      & Error (mm)               & $0.11 \pm 0.23$       & $\mathbf{0.05 \pm 0.02}$       \\
                      & Attempts               & $2.3 \pm 1.3^{*}$           & $\mathbf{1.2 \pm 0.6}^{*}$           \\
\addlinespace
Lateral Ventricles                    & Time (s)                & $\mathbf{87.42 \pm 96.92}$         & $90.70 \pm 49.13$         \\
(Shape IV)  & Error (mm)                & $1.31 \pm 4.08$       & $\mathbf{0.51 \pm 0.32}$       \\
                      & Attempts               & $1.8 \pm 1.8^{*}$           & $\mathbf{1.0 \pm 0.2}^{*}$           \\
\addlinespace
Artery                & Time (s)                & $103.73 \pm 86.68$        & $\mathbf{78.71 \pm 55.23}$         \\
(Shape V)                      & Error (mm)                & $0.48 \pm 0.44$       & $\mathbf{0.31 \pm 0.18}$       \\
                      & Attempts               & $2.8 \pm 2.4^{*}$           & $\mathbf{1.1 \pm 0.3}^{*}$           \\
\addlinespace
Hippocampus           & Time (s)                & $93.32 \pm 60.07$         & $\mathbf{75.98 \pm 55.07}$         \\
(Shape VI)                   & Error (mm)                & $0.32 \pm 0.22$       & $\mathbf{0.22 \pm 0.18}$       \\
                      & Attempts               & $3.0 \pm 3.1^{*}$           & $\mathbf{1.2 \pm 0.5}^{*}$           \\
\bottomrule
\end{tabular}
\end{table}

\subsection{DBS trajectory planning}

\begin{figure}
    \centering
    \includegraphics[width=1\linewidth]{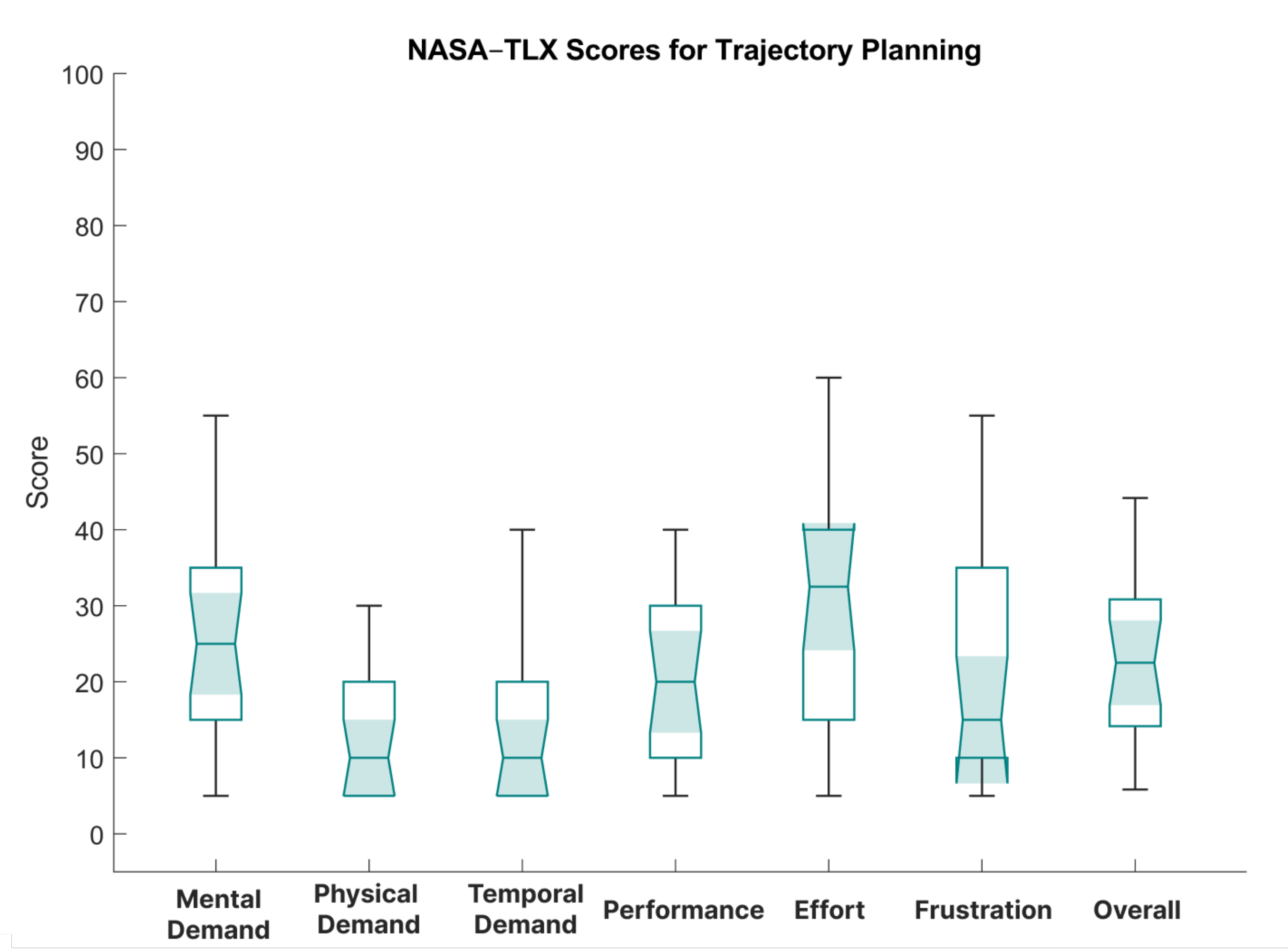}
    \caption{Boxplot of NASA-TLX scores for the proposed mobile AR DBS trajectory planning.}
    \label{fig:NASA-DBS}
\end{figure}

A total of 22 participants completed Phase II: DBS trajectory planning study. \textbf{SUS evaluation} yielded a score of $73.30 \pm 11.45$, significantly above the standard benchmark of 68 ($p = 0.0209$), indicating acceptable overall usability of the mobile AR ``point‐and‐shoot” interface. Perceived task load was further evaluated with \textbf{NASA‐TLX} (Fig. \ref{fig:NASA-DBS}); sub-scale scores (mean$\pm$SD) were: Mental Demand $27.73 \pm 17.57$, Physical Demand $15.45 \pm 14.71$, Temporal Demand $17.27 \pm 15.25$, Performance $25.45 \pm 22.25$, Effort $30.45 \pm 16.32$, Frustration $22.27 \pm 16.74$, and Overall $23.11 \pm 11.38$. These values reflect low perceived task load across all dimensions. Table \ref{tab:stn_placement} summarizes the \textbf{STN surgical target placement metrics} (completion time, targeting error, and number of attempts) for the left and right STN, while Table \ref{tab:stn_trajectory} reports the \textbf{trajectory planning metrics}, including planning time, number of attempts, and distances to vital structures (vessels, ventricles, and sulci). Overall, the targeting error within the STN is lower than 0.5 mm, which is typically desired for DBS procedures, and generally only one attempt is needed. For the trajectory planning, all participants were able to produce valid surgical plans, with just 1$\sim$3 attempts. With the consideration of task completion times for both steps, our proposed system will only require about 2 min to complete a viable surgical trajectory.

\begin{table}[ht]
\centering
\caption{STN Target Placement Metrics (Mean $\pm$ SD).}
\label{tab:stn_placement}
\begin{tabular}{lcc}
\toprule
\textbf{Metric}    & \textbf{Left STN}            & \textbf{Right STN}           \\
\midrule
Time (s)           & $75.30 \pm 41.51$           & $57.79 \pm 27.97$           \\
Error (mm)           & $0.32 \pm 0.19$          & $0.40 \pm 0.12$            \\
Attempts           & $1.1 \pm 0.2$           & $1.0 \pm 0.0$            \\
\bottomrule
\end{tabular}
\end{table}

\vspace{-0.8cm}

\begin{table}[ht]
\centering
\caption{STN Trajectory Metrics (Mean$\pm$SD).}
\label{tab:stn_trajectory}
\begin{tabular}{lcc}
\toprule
\textbf{Metric}                & \textbf{Left STN}            & \textbf{Right STN}           \\
\midrule
Time (s)                       & $70.59 \pm 45.40$           & $60.59 \pm 38.83$           \\
Attempts                       & $1.4 \pm 0.9$           & $1.7 \pm 1.4$            \\
Blood vessel distance (mm)     & $4.64 \pm 1.05$            & $2.79 \pm 0.27$            \\
Ventricle distance (mm)        & $6.21 \pm 2.02$            & $6.27 \pm 3.85$            \\
Sulci distance (mm)            & $2.89 \pm 0.53$            & $2.60 \pm 0.55$            \\
\bottomrule
\end{tabular}
\end{table}

Although our clinical cohort was too small to test between-group differences for statistical significance, clinicians \textit{on average} rated the system as more usable (SUS = 79.50 ± 7.79 vs. 71.47 ± 11.89), took slightly longer on the task (task completion time on both sides = 271.36 ± 114.21 s vs. 262.19 ± 75.75 s), and achieved greater clearance from the sulci (2.75 ± 0.67 mm vs. 2.56 ± 0.52 mm), but not the other critical structures. For STN targeting, the clinicians achieved a similar accuracy (0.35 ± 0.11 mm) to the non-clinicians (0.36 ± 0.17 mm). When looking at the total NASA-TLX score, the clinicians perceived lower task load than the non-clinicians (15.83 ± 4.41 vs. 25.25 ± 11.99). 

\subsection{Qualitative feedback results}

In \textbf{Phase I} study, we further inquired the participants for qualitative feedback on their preference and pros/cons of the two point placement techniques. Among them, 24 out of 25 participants preferred the ``two-line" method, with the most common reasons being ``accuracy or precision" (13/25 users),  ``easier to use" (11/25 users), and ``intuitive" (5/25 users). The one participant that preferred the ``single-line" method reported it was ``easier to use", but also acknowledged that the ``two-line" method is ``more accurate". The primary advantages cited for the ``single-line" method are that it is ``simpler" (10/25 users) and ``faster" (9/25 users). Common disadvantages for the single-line are ``being inaccurate" (10/25 users), ``needing guess work" (6/25 users), and ``higher attempts" (5/25 users). Participants were also asked to suggest improvements. The most common suggestion is to ``improve translation and scaling interaction" (20/25 users) as they both require two-finger interaction and can affect the operation precision of each other at times. To this end, some participants suggested including a toggle between scaling and translation and different finger (gesture) requirements for translation and scaling. Finally, two users also suggested improving the ``single-line" method by allowing further adjustment of point position along the line.

The \textbf{Phase II} study prompted 22 participants for the advantages and disadvantages of the full DBS surgical planning workflow and suggestions for improvements. Generally, the proposed software application is well received, with the participants commenting on it being ``intuitive" (8/22 users), ``visual clarity" (7/22 users), ``easy to use" (5/22 users), and ``confident in selection" (3/22 users). One clinician explicitly stated that the workflow is ``helpful for planning". Similar to Phase I, the most common issue reported by the participants involves the virtual object manipulation with finger gestures (10/22 users). Furthermore, three participants suggested the inclusion of additional anatomical information, such as the skull and cortical parcellations. One participant also suggested to make the surgical target visible regardless of the occlusion to locate it more easily.

\vspace{-0.4cm}

\section{Discussion}\label{sec5}
To date, only a few \cite{Mehta2024, Asadi2025} have explored the use of mobile AR for surgical planning, despite the benefits of its practicality, lower cost, and accessibility. The novelty of our proposed system lies in both the principles of mobile AR user-interaction and the domain of neurosurgical planning. First, we propose efficient and robust ``Point-and-shoot" interaction techniques for both 3D mid-air point placement and constraint-aware trajectory design while tackling missing stereo-vision and spatially tracked annotation tools (e.g., controller or 3D hand tracking) in mobile AR. Although also relying on the raycasting paradigm, prior related methods are tailored for VR headsets \cite{Chen2023,Zhang2022ConductorIB} and are mainly for selecting/placing existing objects on the surface \cite{goh20193d}, which is distinct from our tasks of precision-driven, within-volume 3D surgical target placement \& spatial path planning. Second, compared with other desktop- and VR-based neurosurgical planning software, our proposed system has three main advantages: 1) with just $\sim$2 min/trajectory planning time, it only costs 5\% of the time when using the conventional desktop-based neuronavigation system ($\sim$ 40 min \cite{Beriault2011}). 2) with the use of the more ubiquitous tablet devices, the system can be more accessible in the clinic, particularly for remote areas. 3) in contrast to typical XR headsets, the space requirement is much lower with our system, considering space is often limited in the clinical environment. Additionally, as high accuracy in surgical targeting is critical in functional neurosurgery because small deviations from the intended target can affect therapeutic efficacy \cite{Kremer236}, the sub-millimeter accuracy of our mobile AR system meets the real clinical requirement for DBS procedures. In summary, its high intuitiveness, precision and efficiency directly address the key issues in conventional neuronavigation software, potentially reducing surgical planning time, surgical complications (e.g., intracranial hemorrhage), and operating time. In addition, our approach could complement existing stereotactic workflows and extend to other image-guided neurosurgical procedures requiring precise targeting and surgical trajectory planning.

In the proposed system, the surgical target and trajectory confirmations are achieved by tapping a confirmation button on the bottom right side of the screen when the aiming point is at the desired position (see Fig. 1B). This allows more precise control than directly tapping the desired target spot with a finger, which occludes the view and has a lower precision due to a bigger footprint on the screen. Compared with the proposed spatial AR interaction techniques, conventional 2D annotation by screen-tapping on re-sliced/projected images fails to address the need for 3D spatial interaction in mobile AR surgical planning.

When comparing the proposed ``single-line" vs. ``two-line" point placement methods, the latter is shown to be more advantageous for asymmetric and complex shapes, with reduced task load and significantly fewer attempts for annotation. With the ``single-line" approach, the user must position the mesh model in a strategic pose so that the desired point has equal distance to the front and back of the mesh model along the raycasting direction. Due to the lack of stereovision in the tablet display, it is a bit difficult to achieve the optimal model positioning. With the simpler spheres (Shapes I and II), the ``single-line" technique is quicker and potentially more convenient. However, this is rarely the case for anatomical structures, and as commented by the participants, the ``two-line" method becomes more accurate and easier to use. Note that the reason why we didn't include phantom guides for centroids of the spheres and the STN (in Phase II) is because they are more intuitive to annotate by the non-clinical participants and can help test the capacity of mobile AR for spatial perception. With more complex shapes, especially anatomical models, the phantom guides facilitate the identification of the intended target location for these participants, without expert anatomical knowledge. Aside from the AR view for enhanced spatial understanding \cite{Moro2017}, the ability to scale the model larger can also contribute to the precision of point placement \cite{Hellum2022Voodoo}. For our design, we focus on using touch-based interaction for mid-air annotation while the use of the printed card to anchor the virtual anatomical model also allows additional tablet/card motion-based object manipulation, which could also enhance user experience and outcomes \cite{song2023leveraging}. However, due to lighting conditions and limited display size, large tablet motion for fine-grained point annotation can introduce instability. Our point-and-shoot technique and associated touch-based interaction could mitigate the need of such large device movements. 

More recently, brain tractography \cite{Calabrese2016} is increasingly integrated into the exploration and determination of the optimal stimulation loci within the STN with further personalized symptom improvements. While traditional approach suggests the dorsolateral region of the STN as the initial stimulation target in pre-surgical planning with later intra-operative adjustment based on trial-and-error micro-electrode recordings, individual variability still exists. In this case, modern tractography method that reveals individual patient's brain connectivity could help better address this with interactive exploration in the planning stage. For our first prototype, we use the centroid of STN as a simplistic target to test the system and measure the perceptual accuracy. Future work will further integrate tractography data into the mobile AR system. 

So far, evidence of practical benefits regarding mobile AR has been demonstrated for several surgical applications in recent studies, including superficial circumflex iliac artery perforator flap planning \cite{pereira2019augmented}, cardiovascular surgical planning \cite{Mehta2024} and ventriculostomy planning/guidance \cite{Asadi2025}. These mobile AR applications in surgical planning enable reduced surgical time \cite{pereira2019augmented}, multi-user collaboration \cite{Mehta2024}, and enhanced surgical accuracy \cite{Asadi2025}, with ease of deployment and low cost. These benefits are transferrable to DBS planning, where a high-quality surgical plan is critical in patient outcomes and safety. Our ``Point-and-shoot" methods for mid-air point annotation and surgical trajectory design are proposed based on the ray-casting framework, due to its success in efficient and accurate spatial target selection \cite{Chen2023} and trajectory design \cite{Hellum2022Voodoo, Spiegler2025} in previous VR/AR applications. However, our novel solution tackles the challenge of mid-air point placement inside complex anatomical models in mobile AR for the first time. Designed with a senior neurosurgeon, the proposed mobile AR system for DBS surgical planning is well received by both our clinical and non-clinical participants, yielding a mean SUS of 73.3 (significantly above 68) and low NASA‑TLX scores, with sub‑millimeter surgical target placement error, near‐single‐attempt accuracy, high planning efficiency ($\sim$2 min), and all trajectories meeting the $>$ 2 mm safety margin. These evidences confirm the feasibility and practicality of mobile AR in high-precision surgical planning, such as DBS. Interestingly, the clinicians rated the system more usable and with lower task load than non‐clinicians, likely reflecting their domain expertise in brain anatomy and DBS planning. Compared with the Phase I study, the reduced NASA-TLX score in Phase II may also stem from participants’ prior exposure in Phase I with the same users. 

As the first mobile AR prototype of its kind for DBS planning, there are still a few limitations in our study. \textit{First}, due to the limited access to clinicians with experience in DBS (n=5), we were unable to perform meaningful statistical tests to compare group-wise differences between the clinical and non-clinical participants. However, besides the lay participants, the positive feedback from the domain experts helps confirm the value of the proposed system in this preliminary investigation. In the future, we intend to expand the user study with more clinicians to better understand their needs and further identify improvements required. \textit{Second}, one common issue for the proposed user interface lies in the finger gesture designs for scaling vs translation. As both rely on two fingers, the actions to perform each individual spatial transformations can interfere with each other at times. This makes it more challenging to precisely align the target location with the aiming point on the screen, resulting in extended task completion time. As the two-finger translation and scaling are staples on tablet applications, we didn't make further changes between Phase I and Phase II studies. However, with precision-driven tasks, we noticed that after practicing the interaction in Phase I, the participants perceived the task load to be lower and fewer reported the translation vs. scaling issue in Phase II. In the next iteration of the system, we will further incorporate the improvement suggestions from the user study, with a focus on better precision control for the finger-gesture-based 3D model manipulation. \textit{Finally}, our current user study is still limited by the session duration due to participant fatigue and clinical experts' busy schedules, as well as accessibility of the limited prior XR surgical planning software. While we have included a comprehensive list of qualitative and quantitative metrics that are consistent with prior similar publications and clinical standards for STN DBS planning \cite{Hellum2022Voodoo,Spiegler2025,Beriault2011} in both phases of our study, it is still desirable to further include additional variations of brain models, broader evaluations across procedures (e.g, SEEG, EVD, and cerebrovascular surgery), and within-participant comparison of other XR and commercial surgical planning solutions in future investigations.

\vspace{-0.5cm}

\section{Conclusion}
In this study, we introduce a novel tablet-based mobile AR application for neurosurgical trajectory planning, with application to DBS procedures. Specifically, we designed and compared two novel interaction strategies to place 3D surgical target points within anatomical structures, and confirmed the superiority of the ``two-line" ray-shooting approach. With a ``point-and-shoot" interaction strategy, our proposed AR system enables efficient and robust surgical trajectory planning, with good usability as demonstrated through a user study. As a more accessible and portable solution for navigating complex 3D anatomical data, we will incorporate the feedback from the study and further improve the system, with an expanded study in additional surgical applications that include more clinicians in the future. 

% %\backmatter
\bmsubsection*{Author Contributions}
\textbf{Li Wah D’Angelo}: Conceptualization; formal analysis; investigation; methodology; software; visualization; writing--original draft. 
\textbf{Pascal Spiegler}: formal analysis; visualization; writing--original draft.  
\textbf{Aristides Hadjinicolaou}: validation; writing—review and editing. 
\textbf{Alexander G. Weil}: Methodology; validation; writing—review and editing.
\textbf{Yiming Xiao}: Conceptualization; formal analysis; methodology; supervision;
visualization; writing—review and editing.

%\bmsubsection*{Acknowledgments}
%xxxxxx

\bmsubsection*{Conflicts of Interest}
The authors declare no conflicts of interest.

%\bmsubsection*{Data availability statement}
%Data sharing not applicable no new data generated, or the article describes entirely theoretical research.

\bibliography{template}

\begin{thebibliography}{34}
\providecommand{\natexlab}[1]{#1}
\providecommand{\url}[1]{\texttt{#1}}
\expandafter\ifx\csname urlstyle\endcsname\relax
  \providecommand{\doi}[1]{doi: #1}\else
  \providecommand{\doi}{doi: \begingroup \urlstyle{rm}\Url}\fi

\bibitem[\protect\citeauthoryear{Xiao}{Xiao}{}]{xiao2021frqnt}
Yiming Xiao 2021.
\newblock ``Vers un système de visualisation intelligent et intuitif pour la neurochirurgie.''   \url{https://doi.org/10.69777/296459}, Projets de recherche en équipe, Fonds de recherche du Québec (FRQ).

\bibitem[\protect\citeauthoryear{Xiao}{Xiao}{}]{xiao2024frq}
Yiming Xiao 2024.
\newblock ``Améliorer le prognostic neurochirurgical avec l'imagerie multimodale et à la connectivité cérébrale.''   \url{https://doi.org/10.69777/330745}, Chercheurs-boursiers, numéro de demande 330745, Fonds de recherche du Québec (FRQ).

\bibitem[\protect\citeauthoryear{Bériault, Subaie, Collins, Sadikot, and Pike}{Bériault et~al.}{}]{Beriault2012MultiModal}
S.~Bériault, F.~A. Subaie, D.~L. Collins, A.~F. Sadikot,, and G.~B. Pike.
\newblock
\newblock ``A multi-modal approach to computer-assisted deep brain stimulation trajectory planning.''  {\it International Journal of Computer Assisted Radiology and Surgery\/}~7, no. 5 (2012): 687--704.

\bibitem[\protect\citeauthoryear{Beriault, Al~Subaie, Mok, Sadikot, and Pike}{Beriault et~al.}{}]{Beriault2011}
S.~Beriault, F.~Al~Subaie, K.~Mok, A.~F. Sadikot,, and G.~B. Pike.
\newblock
\newblock ``Automatic trajectory planning of DBS neurosurgery from multi-modal MRI datasets.''  {\it Med Image Comput Comput Assist Interv\/}~14, no. Pt 1 (2011): 259--66.

\bibitem[\protect\citeauthoryear{Zelmann, Beriault, Marinho, Mok, Hall, Guizard, Haegelen, Olivier, Pike, and Collins}{Zelmann et~al.}{}]{Zelmann2015Improving}
R.~Zelmann, S.~Beriault, M.~M. Marinho, K.~Mok, J.~A. Hall, N.~Guizard, C.~Haegelen, A.~Olivier, G.~B. Pike,, and D.~L. Collins.
\newblock
\newblock ``Improving recorded volume in mesial temporal lobe by optimizing stereotactic intracranial electrode implantation planning.''  {\it International Journal of Computer Assisted Radiology and Surgery\/}~10, no. 10 (2015): 1599--1615.

\bibitem[\protect\citeauthoryear{Tene, Vique~L{\'o}pez, Valverde~Aguirre, Orna~Puente, and Vacacela~Gomez}{Tene et~al.}{}]{tene2024virtual}
Talia Tene, Diego~Fabi{\'a}n Vique~L{\'o}pez, Paulina~Elizabeth Valverde~Aguirre, Luz~Mar{\'\i}a Orna~Puente,, and Cristian Vacacela~Gomez.
\newblock
\newblock ``Virtual reality and augmented reality in medical education: an umbrella review.''  {\it Frontiers in Digital Health\/}6 (2024): 1365345.

\bibitem[\protect\citeauthoryear{Pinter, Lasso, Choueib, Asselin, Fillion-Robin, Vimort, Martin, Jolley, and Fichtinger}{Pinter et~al.}{}]{pinter2020slicervr}
Csaba Pinter, Andras Lasso, Saleh Choueib, Mark Asselin, Jean-Christophe Fillion-Robin, Jean-Baptiste Vimort, Ken Martin, Matthew~A Jolley,, and Gabor Fichtinger.
\newblock
\newblock ``SlicerVR for medical intervention training and planning in immersive virtual reality.''  {\it IEEE transactions on medical robotics and bionics\/}~2, no. 2 (2020): 108--117.

\bibitem[\protect\citeauthoryear{Hellum, Steele, and Xiao}{Hellum et~al.}{}]{Sonia}
Owen Hellum, Christopher Steele,, and Yiming Xiao.
\newblock
\newblock ``SONIA: an immersive customizable virtual reality system for the education and exploration of brain networks.''  {\it Frontiers in Virtual Reality\/}4.

\bibitem[\protect\citeauthoryear{Hellum, Mu, Kersten-Oertel, and Xiao}{Hellum et~al.}{}]{Hellum2022Voodoo}
Owen Hellum, Yanyu Mu, Marta Kersten-Oertel,, and Yiming Xiao.
\newblock
\newblock ``A novel prototype for virtual-reality-based deep brain stimulation trajectory planning using voodoo doll annotation and eye-tracking.''  {\it Computer Methods in Biomechanics and Biomedical Engineering: Imaging \& Visualization\/}~10, no. 4 (2022): 418--424.

\bibitem[\protect\citeauthoryear{Harirpoush, Rakovich, Kersten-Oertel, and Xiao}{Harirpoush et~al.}{}]{Harirpoush2024}
Arash Harirpoush, George Rakovich, Marta Kersten-Oertel,, and Yiming Xiao.
\newblock
\newblock ``Virtual reality-based preoperative planning for optimized trocar placement in thoracic surgery: A preliminary study.''  {\it Healthcare Technology Letters\/}~11, no. 6 (2024): 418--426.

\bibitem[\protect\citeauthoryear{Spiegler, Abdelsalam, Hellum, Hadjinicolaou, Weil, and Xiao}{Spiegler et~al.}{}]{Spiegler2025}
Pascal Spiegler, Haitham Abdelsalam, Owen Hellum, Aristides Hadjinicolaou, Alexander~G. Weil,, and Yiming Xiao.
\newblock
\newblock ``PreVISE: an efficient virtual reality system for SEEG surgical planning.''  {\it Virtual Reality\/}~29, no. 1 (2024): 13.

\bibitem[\protect\citeauthoryear{Fotouhi, Mehrfard, Song, Johnson, Osgood, Unberath, Armand, and Navab}{Fotouhi et~al.}{}]{fotouhi2020development}
Javad Fotouhi, Arian Mehrfard, Tianyu Song, Alex Johnson, Greg Osgood, Mathias Unberath, Mehran Armand,, and Nassir Navab.
\newblock
\newblock ``Development and pre-clinical analysis of spatiotemporal-aware augmented reality in orthopedic interventions.''  {\it IEEE transactions on medical imaging\/}~40, no. 2 (2020): 765--778.

\bibitem[\protect\citeauthoryear{Moglia, Marsilio, Rossi, Pinelli, Lettieri, Mainardi, Manzotti, and Cerveri}{Moglia et~al.}{}]{moglia2023mixed}
Andrea Moglia, Luca Marsilio, Matteo Rossi, Maria Pinelli, Emanuele Lettieri, Luca Mainardi, Alfonso Manzotti,, and Pietro Cerveri.
\newblock
\newblock ``Mixed reality and artificial intelligence: A holistic approach to multimodal visualization and extended interaction in knee osteotomy.''  {\it IEEE Journal of Translational Engineering in Health and Medicine\/}12 (2023): 279--290.

\bibitem[\protect\citeauthoryear{Kunz, Hlav{\'a}{\v{c}}, Schneider, Pala, Henrich, Jickeli, W{\"o}rn, Hein, Wirtz, and Mathis-Ullrich}{Kunz et~al.}{}]{kunz2021autonomous}
Christian Kunz, Michal Hlav{\'a}{\v{c}}, Max Schneider, Andrej Pala, Pit Henrich, Birgit Jickeli, Heinz W{\"o}rn, Bj{\"o}rn Hein, Rainer Wirtz,, and Franziska Mathis-Ullrich.
\newblock
\newblock ``Autonomous planning and intraoperative augmented reality navigation for neurosurgery.''  {\it IEEE Transactions on Medical Robotics and Bionics\/}~3, no. 3 (2021): 738--749.

\bibitem[\protect\citeauthoryear{De~Jesus Encarnacion~Ramirez, Chmutin, Nurmukhametov, Soto, Kannan, Piavchenko, Nikolenko, Efe, Romero, Mukengeshay, et~al.}{De~Jesus Encarnacion~Ramirez et~al.}{}]{de2024integrating}
Manuel De~Jesus Encarnacion~Ramirez, Gennady Chmutin, Renat Nurmukhametov, Gervith~Reyes Soto, Siddarth Kannan, Gennadi Piavchenko, Vladmir Nikolenko, Ibrahim~E Efe, Alberto~Ram{\'\i}rez Romero, Jeff~Ntalaja Mukengeshay, et~al.
\newblock
\newblock ``Integrating augmented reality in spine surgery: redefining precision with new technologies.''  {\it Brain Sciences\/}~14, no. 7 (2024): 645.

\bibitem[\protect\citeauthoryear{Mehta, Karanth, Yang, Slesnick, Shaw, and Chau}{Mehta et~al.}{}]{Mehta2024}
Pratham~Darrpan Mehta, Harsha Karanth, Haoyang Yang, Timothy~C Slesnick, Fawwaz Shaw,, and Duen~Horng Chau. 2024.
\newblock ``ARCollab: Towards Multi-User Interactive Cardiovascular Surgical Planning in Mobile Augmented Reality.''   In {\it Extended Abstracts of the CHI Conference on Human Factors in Computing Systems}, CHI EA '24, New York, NY, USA: Association for Computing Machinery.

\bibitem[\protect\citeauthoryear{Asadi, Castillo, Asadi, Sinclair, and Kersten-Oertel}{Asadi et~al.}{}]{Asadi2025}
Zahra Asadi, Joshua~Pardillo Castillo, Mehrdad Asadi, David~S. Sinclair,, and Marta Kersten-Oertel.
\newblock
\newblock ``iSurgARy: A mobile augmented reality solution for ventriculostomy in resource-limited settings.''  {\it Healthcare Technology Letters\/}~12, no. 1 (2025): e12118.

\bibitem[\protect\citeauthoryear{Hellum, Kersten-Oertel, and Xiao}{Hellum et~al.}{}]{Hellum2023}
Owen Hellum, Marta Kersten-Oertel,, and Yiming Xiao.
\newblock
\newblock ``Assessment of user-interaction strategies for neurosurgical data navigation and annotation in virtual reality.''  {\it Virtual Reality\/}~27, no. 2 (2023): 1345--1355.

\bibitem[\protect\citeauthoryear{Goh, Sunar, and Ismail}{Goh et~al.}{}]{goh20193d}
Eg~Su Goh, Mohd~Shahrizal Sunar,, and Ajune~Wanis Ismail.
\newblock
\newblock ``3D object manipulation techniques in handheld mobile augmented reality interface: A review.''  {\it IEEE Access\/}7 (2019): 40581--40601.

\bibitem[\protect\citeauthoryear{Polvi, Taketomi, Yamamoto, Dey, Sandor, and Kato}{Polvi et~al.}{}]{polvi2016slidar}
Jarkko Polvi, Takafumi Taketomi, Goshiro Yamamoto, Arindam Dey, Christian Sandor,, and Hirokazu Kato.
\newblock
\newblock ``SlidAR: A 3D positioning method for SLAM-based handheld augmented reality.''  {\it Computers \& Graphics\/}55 (2016): 33--43.

\bibitem[\protect\citeauthoryear{Song, Eck, and Navab}{Song et~al.}{}]{song2023leveraging}
Tianyu Song, Ulrich Eck,, and Nassir Navab. 2023.
\newblock ``Leveraging Motion Tracking for Intuitive Interactions in a Tablet-Based 3D Scene Annotation System.''   In {\it 2023 IEEE International Symposium on Mixed and Augmented Reality (ISMAR)},   563--572.
\newblock IEEE.

\bibitem[\protect\citeauthoryear{Grandi, Debarba, Bemdt, Nedel, and Maciel}{Grandi et~al.}{}]{grandi2018design}
Jer{\^o}nimo~G Grandi, Henrique~G Debarba, Iago Bemdt, Luciana Nedel,, and Anderson Maciel. 2018.
\newblock ``Design and assessment of a collaborative 3D interaction technique for handheld augmented reality.''   In {\it 2018 IEEE Conference on Virtual Reality and 3D User Interfaces (VR)},   49--56.
\newblock IEEE.

\bibitem[\protect\citeauthoryear{Xiao, Lau, Anderson, DeKraker, Collins, Peters, and Khan}{Xiao et~al.}{}]{Xiao2019}
Yiming Xiao, Jonathan~C. Lau, Taylor Anderson, Jordan DeKraker, D.~Louis Collins, Terry Peters,, and Ali~R. Khan.
\newblock
\newblock ``An accurate registration of the BigBrain dataset with the MNI PD25 and ICBM152 atlases.''  {\it Scientific Data\/}~6, no. 1 (2019): 210.

\bibitem[\protect\citeauthoryear{Yang, Xia, Kin, and Igarashi}{Yang et~al.}{}]{yang2020intra}
Xi~Yang, Ding Xia, Taichi Kin,, and Takeo Igarashi. 2020.
\newblock ``IntrA: 3D Intracranial Aneurysm Dataset for Deep Learning.''   In {\it The IEEE Conference on Computer Vision and Pattern Recognition (CVPR)}, .

\bibitem[\protect\citeauthoryear{Frangi, Niessen, Vincken, and Viergever}{Frangi et~al.}{}]{Frangi}
Alejandro~F. Frangi, Wiro~J. Niessen, Koen~L. Vincken,, and Max~A. Viergever. 1998.
\newblock ``Multiscale vessel enhancement filtering.''   In {\it Medical Image Computing and Computer-Assisted Intervention — MICCAI’98}, edited by William~M. Wells, Alan Colchester, and Scott Delp,   130--137.
\newblock Springer Berlin Heidelberg.

\bibitem[\protect\citeauthoryear{Eskildsen, Coupe, Fonov, Manjon, Leung, Guizard, Wassef, Ostergaard, Collins, and Alzheimer's Disease~Neuroimaging}{Eskildsen et~al.}{}]{BEAST}
S.~F. Eskildsen, P.~Coupe, V.~Fonov, J.~V. Manjon, K.~K. Leung, N.~Guizard, S.~N. Wassef, L.~R. Ostergaard, D.~L. Collins,, and Initiative Alzheimer's Disease~Neuroimaging.
\newblock
\newblock ``BEaST: brain extraction based on nonlocal segmentation technique.''  {\it Neuroimage\/}~59, no. 3 (2012): 2362--73.

\bibitem[\protect\citeauthoryear{Hart}{Hart}{}]{NASATLX}
Sandra~G. Hart.
\newblock
\newblock ``Nasa-Task Load Index (NASA-TLX); 20 Years Later.''  {\it Proceedings of the Human Factors and Ergonomics Society Annual Meeting\/}~50, no. 9 (2006): 904--908.

\bibitem[\protect\citeauthoryear{Brooke}{Brooke}{}]{SUS68}
John Brooke.
\newblock
\newblock ``SUS: a retrospective.''  {\it Journal of Usability Studies\/}8 (2013): 29--40.

\bibitem[\protect\citeauthoryear{Chen, Giordano, Benko, Grossman, and Santosa}{Chen et~al.}{}]{Chen2023}
Di~Laura Chen, Marcello Giordano, Hrvoje Benko, Tovi Grossman,, and Stephanie Santosa. 2023.
\newblock ``GazeRayCursor: Facilitating Virtual Reality Target Selection by Blending Gaze and Controller Raycasting.''   In {\it Proceedings of the 29th ACM Symposium on Virtual Reality Software and Technology}, VRST '23, New York, NY, USA: Association for Computing Machinery.

\bibitem[\protect\citeauthoryear{Zhang, Katsuragawa, and Lank}{Zhang et~al.}{}]{Zhang2022ConductorIB}
Futian Zhang, Keiko Katsuragawa,, and Edward Lank.
\newblock
\newblock ``Conductor: Intersection-Based Bimanual Pointing in Augmented and Virtual Reality.''  {\it Proceedings of the ACM on Human-Computer Interaction\/}6 (2022): 103 -- 117.

\bibitem[\protect\citeauthoryear{Kremer, van Laar, Lange, Statius~Muller, la~Bastide-van Gemert, Oterdoom, Drost, and van Dijk}{Kremer et~al.}{}]{Kremer236}
Naomi~I Kremer, Teus van Laar, St{\`e}fan~F Lange, Sijmen Statius~Muller, Sacha la~Bastide-van Gemert, DL~Marinus Oterdoom, Gea Drost,, and J~Marc~C van Dijk.
\newblock
\newblock ``STN-DBS electrode placement accuracy and motor improvement in Parkinson{\textquoteright}s disease: systematic review and individual patient meta-analysis.''  {\it Journal of Neurology, Neurosurgery \& Psychiatry\/}~94, no. 3 (2023): 236--244.

\bibitem[\protect\citeauthoryear{Moro, Stromberga, Raikos, and Stirling}{Moro et~al.}{}]{Moro2017}
C.~Moro, Z.~Stromberga, A.~Raikos,, and A.~Stirling.
\newblock
\newblock ``The effectiveness of virtual and augmented reality in health sciences and medical anatomy.''  {\it Anat Sci Educ\/}~10, no. 6 (2017): 549--559.

\bibitem[\protect\citeauthoryear{Calabrese}{Calabrese}{}]{Calabrese2016}
E.~Calabrese.
\newblock
\newblock ``Diffusion Tractography in Deep Brain Stimulation Surgery: A Review.''  {\it Front Neuroanat\/}10 (2016): 45.

\bibitem[\protect\citeauthoryear{Pereira, Kufeke, Parada, Troncoso, Bahamondes, Sanchez, and Roa}{Pereira et~al.}{}]{pereira2019augmented}
Nicol{\'a}s Pereira, Mat{\'\i}as Kufeke, Leonardo Parada, Ekaterina Troncoso, Jorge Bahamondes, Luis Sanchez,, and Ricardo Roa.
\newblock
\newblock ``Augmented reality microsurgical planning with a smartphone (ARM-PS): a dissection route map in your pocket.''  {\it Journal of Plastic, Reconstructive \& Aesthetic Surgery\/}~72, no. 5 (2019): 759--762.

\end{thebibliography}

% \bmsubsection*{Supporting Information}

% Additional supporting information can be found online in the Supporting Information
% section.

\end{document}